# Cascaded multiplexed optical link on a telecommunication network for frequency dissemination

Olivier Lopez[1], Adil Haboucha[2], Fabien Kéfélian[1], Haifeng Jiang[2], Bruno Chanteau[1], Vincent Roncin[1], Christian Chardonnet[1], Anne Amy-Klein[1*] and Giorgio Santarelli[2]

[1]*Laboratoire de Physique des Lasers, Université Paris 13, CNRS, 99 Avenue Jean-Baptiste Clément, 93430 Villetaneuse, France*
[2]*Laboratoire National de Métrologie et d'Essais–Système de Références Temps-Espace, Observatoire de Paris, CNRS, UPMC, 61 Avenue de l'Observatoire, 75014 Paris, France*
[*]*anne.amy-klein@univ-paris13.fr*

**Abstract:** We demonstrate a cascaded optical link for ultrastable frequency dissemination comprised of two compensated links of 150 km and a repeater station. Each link includes 114 km of Internet fiber simultaneously carrying data traffic through a dense wavelength division multiplexing technology, and passes through two routing centers of the telecommunication network. The optical reference signal is inserted in and extracted from the communication network using bidirectional optical add-drop multiplexers. The repeater station operates autonomously ensuring noise compensation on the two links and the ultra-stable signal optical regeneration. The compensated link shows a fractional frequency instability of $3\times10^{-15}$ at one second measurement time and $5\times10^{-20}$ at 20 hours. This work paves the way to a wide dissemination of ultra-stable optical clock signals between distant laboratories via the Internet network.

## 1. Introduction

The transfer of ultra-stable frequency signals between distant laboratories is required by many applications in time and frequency metrology, fundamental physics, particle accelerators, and astrophysics [1-8]. Stable radio and microwave frequency transmission over an optical link has already been demonstrated [2-5, 9-12]. Frequency transfer using the optical phase of an ultra-stable laser over a dedicated fiber link was reported on distances up to about 200 km by several groups [13-19]. The current challenge is to extend this technique of frequency dissemination on longer distances in order to connect laboratories of different countries.

Scalability of the roundtrip noise compensation to long-haul optical link is not straightforward: attenuation and phase noise increase with the fiber length whereas correction bandwidth and noise rejection decrease with it [13]. The geographic situation of the link has also a strong impact since uncompensated link phase noise power spectral density per unit of length can vary over 3 orders of magnitude [19, 20]. A 900 km dedicated optical link has been demonstrated very recently with a very good resolution thanks to the intrinsic low-noise of the link [21]. For a noisier link, a multiple segments approach can be implemented [15]. The number and length of each segment can be optimized depending on the noise distribution along the link.

We have recently developed a frequency dissemination approach which takes advantage of the existing Internet fiber network already connecting research facilities and universities via the National Research Networks (NRENs) [22]. With this approach, the ultra-stable signal propagates directly on a telecommunication fiber simultaneously transmitting digital data, using dense wavelength division multiplexing (DWDM) technology.

Internet fibers present a higher number of connectors and additional specific WDM equipment, as compared to dedicated fibers, which leads to an increased attenuation and propagation noise. The multiple sub-link approach allows for an increased correction bandwidth and robustness regarding attenuation, but requires the development of a repeater station between each segment.

In this paper, we present the first demonstration of a cascaded optical link for frequency transfer including two segments of internet network fibers and one repeater station. A major part of the link uses telecommunication fibers simultaneously transmitting digital data of the Internet network. In a first part we will describe the repeater station which has to work autonomously in a telecommunication network node. Then we will present the architecture of the cascaded link we have implemented. Finally results will be presented and discussed.

## 2. Repeater station for multi-segment link

### 2.1 Principle of operation

With a multiple segments approach, intermediate stations are required where a low-noise optical oscillator is phase locked to the incoming signal in order to regenerate a spectrally pure signal seeding the next segment. The purpose of the $N^{th}$ station along the link is threefold as depicted in Figure 1. First it sends back part of the incoming optical signal to the previous station (N-1), after amplification and frequency shift, in order to stabilize the incoming link (named link N). Second, the station regenerates the incoming optical signal thanks to a clean-up optical tracking oscillator based on a low-noise Bragg grating fiber laser phase-locked on the incoming signal. The regenerated signal then feeds the link departing to the next station (N+1). Third, it processes the signal sent back from station N+1, allowing for noise compensation of the following link segment (link N+1).

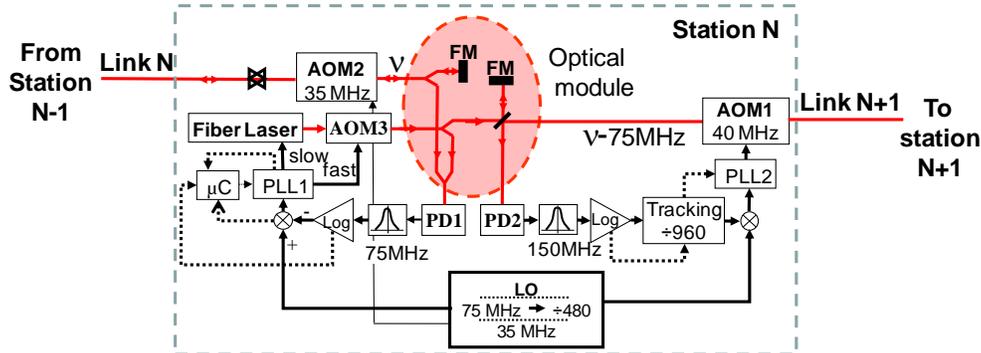

Fig. 1. Scheme of the $N^{th}$ repeater station, FM : Faraday mirror, PD : photodiode, LO : local RF oscillator, AOM : acousto-optic modulator, Log : logarithmic amplifier, µC : microcontroller, PLL : phase-lock loop

The compensation is based on the assumption that the laser phase fluctuations due to propagation are identical in both directions. This usual assumption is valid for symmetrical propagation characteristics (frequency and polarisation state). The use of a Faraday mirror imposes polarisation state retrace of the return beam. The forward and backward signals have a frequency difference of 2x35 MHz. Since the phase is, at first order, proportional to the frequency, it induces different forward and backward phase fluctuations. This leads to an imperfect determination of the phase correction of the link which is proportional to the free running link phase fluctuations multiplied by the ratio between the frequency offset and the optical frequency ($3 \times 10^{-7}$). In principle this error can be compensated by digital frequency processing. However if we assume a long term free running instability of a few $10^{-14}$, this gives an upper limit for the link resolution below $10^{-20}$. Finally the Sagnac effect, which is intrinsically asymmetric, is an ultimate limit for frequency resolution.

To implement this approach, repeater stations are designed to be installed in remote telecommunication sites which are noisier and have greater temperature fluctuations than the laboratory environment. Consequently, a highly robust and reliable system is required. Each station consists of two modules, one for the electronics package and the other for the optical part.

*2.2 Optical set-up*

The optical set-up of each station consists of a laser, an optical module, three acousto-optic modulators (AOM) and two photodiodes (PD). For the local optical oscillator we use a fiber laser system based on a distributed-feedback technique (Koheras Basik OEM). Its free-running linewidth is about 1kHz. Its frequency can be tuned over a range of ~ 2 GHz, which is sufficient for compensating the long term noise of this laser by controlling the PZT port. Indeed a ± 1 K fluctuation of the room temperature corresponds to a variation of the optical frequency smaller than 500 MHz.

The compact optical module has been realized by splicing off-the-shelf fiber-optic components. This module contains all the critical components of the station in term of phase stability : 2 Faraday mirrors, 2 isolators (in front of the photodiodes) and a few couplers. Consequently, the fiber pigtail lengths have been minimized and finely matched in order to reduce the effect of the residual thermal fluctuation of the non-common paths' lengths. The whole optical circuit is housed in an aluminum box (dimensions 30x100x180 mm) covered by a thick polyurethane foam layer (50mm) for acoustic noise and temperature shielding. Moreover the temperature is actively stabilized around 298 K using a Peltier element. The temperature varies less than 20 mK when the ambient temperature fluctuates about 1 K.

The optical set-up is designed to generate two beat-note signals (see Fig. 1). The first one, detected by photodiode 1 (PD1), is used to lock the local laser on the incoming signal from link N using the laser PZT and AOM3. The second beat-note signal on PD2 quantifies the roundtrip phase fluctuations caused by acoustical, mechanical, and thermal delay propagation fluctuations along link N+1. AOM1 is then used to compensate for this fiber induced noise. AOM2 shifts the frequency of the return signal to station N-1 in order to distinguish it from parasitic reflections.

*2.3 Electronic set-up*

The electronic system can be seen on Fig. 1. The beatnote signal for the laser lock is first filtered with a 75 MHz band pass filter to reduce the noise bandwidth to about 10 MHz. In order to make the control system insensitive to amplitude fluctuations, we use a logarithmic amplifier. A digital phase-frequency detector generates the phase-error signal processed by a loop filter (PLL1). Fast corrections (100 kHz bandwidth) are applied through the voltage controlled oscillator driving AOM3. Slow corrections are applied to the laser PZT input with a few tens of Hz bandwidth.

The beatnote signal for the N+1 link stabilization is filtered with a narrow 150 MHz band pass filter (5 MHz bandwidth) to reject the parasitic reflections and amplified with a logarithmic amplifier. A tracking oscillator is then used to filter the signal in an ultra-narrow bandwidth of 100 kHz. The tracking oscillator signal, after a digital division by 960, is mixed with the local RF oscillator signal at 75MHz, itself divided by 480. This signal is processed by PLL2 and applied to AOM1. The division is necessary since the link phase noise is very high and it enables increasing dynamic compensation control, hence improving the robustness of PLL2.

*2.4 The station automatic operation*

The station is automatically operated by microcontrollers in order to achieve autonomous operation. First of all, one microcontroller manages the local laser phase-lock acquisition. The laser frequency is scanned until the beatnote signal at the logarithmic amplifier exceeds a pre-defined power level. This ensures that it is inside the capture range of the phase lock loop. Consequently the loop is closed by the microcontroller, and the laser frequency is locked at the frequency ν-75 MHz, where ν is the received laser frequency. At a second step, the tracking oscillator loop is closed, and finally PLL2 is closed.

All RF signals are synthesized from a single local RF reference (see Fig.1). Thanks to the optimized choice of the beat-note frequencies at 75 and 150 MHz, any possible drift of the local oscillator has no influence on the N to N+1 frequency transfer. The local laser is locked to the input signal (after AOM2) with a -75 MHz shift. The transferred signal to the next station is shifted to this local laser by +75 MHz, since the return signal from the next station is locked to the local laser signal with a 150 MHz shift. Both frequency shifts are synthesized by the same RF local oscillator of the station. Since they are opposite in sign, their sum is exactly zero, whatever the drift of the local oscillator.

## 3. Cascaded Link

### 3.1 Station testbed

To characterize the stability performance of the repeater station, we have used fiber spools in order to mimic the installed fiber span. The testbed link is comprised of two segments of 150 km length cascaded with the repeater station, as described in section 2. Each segment includes 4 optical add drop multiplexers (OADM) and 2 bidirectional Erbium-doped fiber amplifiers (EDFA) as in a real multiplexed link (see below). At the link input, an identical station, denoted as input station, is fed with a cavity-stabilized laser [16]. The link performance is evaluated by measuring the end-to-end beatnote signal between the local laser of the input station and the end of the link. For instability measurements, we use a multichannel zero dead-time counter which operates with a Π-type window. Two channels simultaneously record the frequency of the compensated link with a bandwidth of 100 kHz and 10 Hz respectively. A third channel is used to record the AOM1 frequency, which represents the free-running frequency noise of the first link of 150 km. The fractional Allan deviation is then calculated from the frequency samples and is shown in Fig. 2.

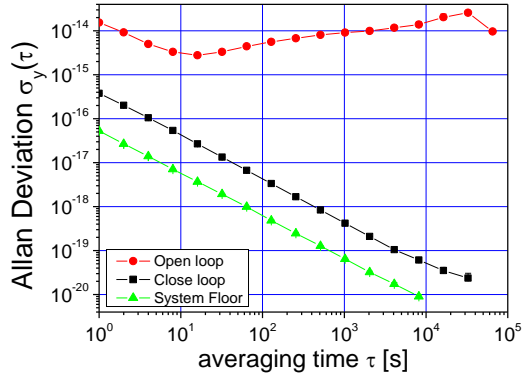

Fig. 2. End-to-end fractional frequency instability of a 2x150 km cascaded testbed link: 150-km free running up-link (red circles), 2x150 km compensated link measured with a 10 Hz filter (black squares), system floor with zero-length link (green up-triangles)

The stability of the testbed link for a 10 Hz bandwidth (black squares) is $3\times10^{-16}$ at 1 s averaging time, averaging down with a $1/\tau$ slope to the mid $10^{-20}$ range at 30000 s. This very good stability arises from the lower noise of the spools (red circles) compared to an installed fiber. On the same graph is plotted the system noise floor of the frequency transfer obtained with a zero-length link (green up-triangles), which reaches the $10^{-20}$ level at $10^4$ s.

*3.2 Cascaded link using the telecommunication network*

Using this station, we have demonstrated a first implementation of a 300-km cascaded optical link on the French academic and research network (RENATER).

The overall scheme of the 300 km-long optical link (LPL-Nogent l'Artaud-LPL) with an intermediate repeater station located at Nogent l'Artaud (100 km east of Paris) is depicted in Fig. 3. This link starts and ends at LPL laboratory (at Université Paris 13, in the north surroundings of Paris), and is composed of three different fiber spans. In each span, there are two identical parallel fibers labeled as UL and DL. UL is used for the uplink consisting of 114 km of fiber carrying internet traffic and 36 km of dedicated dark fiber. DL is used for the downlink with the same characteristics. The first span is composed of two 11 km-long fibers connecting the information service and technology center of Université Paris 13 to a Data Centre Facility (DCF) located in Aubervilliers (Interxion1). The digital stream between Université Paris 13 and Aubervilliers is encoded over an optical carrier on channel #34 (1550.12 nm) of the International Telecommunication Union (ITU) grid whereas the ultrastable signal is carried by the channel #44, at 1542.14 nm. The second span is composed of two 36 km-long urban dark fibers which connect the two DCFs of Interxion 1 and TeleHouse 2, downtown Paris. The third span is composed of two 103 km long-haul intercity fibers simultaneously carrying internet data traffic. In this part the digital data signals are transferred over channels #42 and #43, only 0.8 nm and 1.6 nm away from the ultrastable signal. At Nogent l'Artaud the ultra-stable signal arriving from UL is processed in the repeater station and sent into the other 150 km-long fiber (DL) back to Université Paris 13.

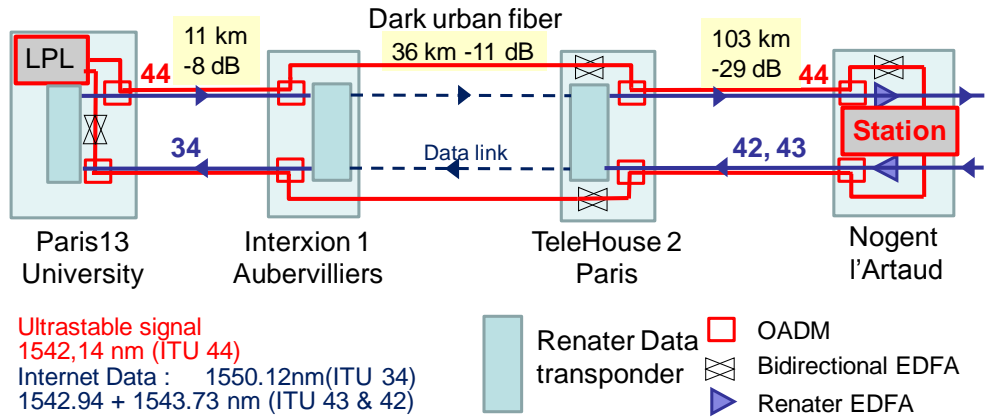

Fig. 3. Scheme of the 2 x 150 km cascaded optical link.

Eight OADM are used to insert and extract the ultrastable signal from the Internet fibers. These three-port components can insert or extract the 1542.14 nm wavelength into or out of the other wavelengths with isolation better than 25 dB for an adjacent channel (100 GHz) and better than 40 dB for other channels. Attenuation is an issue since the round-trip propagation of the optical signal is required over the same fiber for noise compensation. Total roundtrip attenuation for each 150 km link is more than 100 dB due to fiber losses, OADMs and the large number of connectors. In order to overcome these losses, we used two bidirectional EDFA for each link. Better amplification performance could be obtained with Fiber Brillouin Amplifiers as demonstrated in [23].

**4. Results and discussions**

Fig. 4 shows the temporal behavior of the link propagation delay over 3.5 days. The (a) curve displays the 300-km free-running propagation delay measured by replacing the Nogent l'Artaud station with an optical amplifier and using fixed AOM frequency shifts. It gives the free-running fiber propagation delay fluctuation, which spans over a few ns. This corresponds to a temperature mean change of the order of 0.1 K. This is quite low because the fibers are either buried or inside underground service facilities, limiting environment perturbation. The (b) and (c) curves display the cascaded 300-km propagation delay fluctuation, respectively in closed loop and in open loop for the up-link. They were recorded at the same time. The closed loop signal is confined to below 10 fs peak-to-peak, which demonstrates that the correction systems have a rejection factor close to $10^5$. The open loop signal exhibits higher fluctuation than the free-running fiber fluctuation. It corresponds to the fluctuation of the whole up-link, including AOM2, which is driven by the free local oscillator of the repeater station at Nogent l'Artaud.

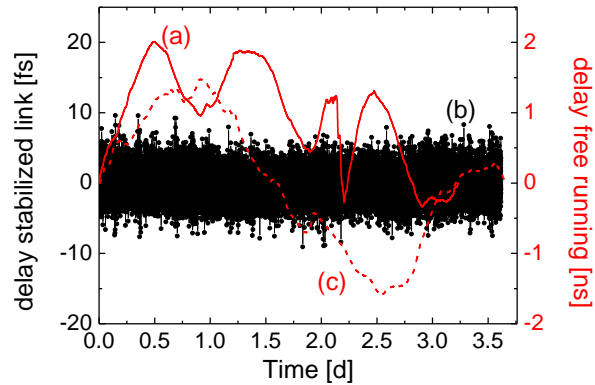

Fig. 4. Temporal behavior (sampled every 10s) of the optical link propagation delay: (a) end-to-end propagation delay of the 300 km free-running optical link recorded without repeater station at Nogent l'Artaud (red trace); (b) end-to-end propagation delay of the 300-km cascaded compensated link (black trace), (c) correction signal (applied to the input AOM at LPL) of the 150-km up-link (red dashed trace). Curves (b) and (c) were recorded simultaneously.

Figure 5 shows the phase noise spectral density of the various segments composing the link. The phase noise is evaluated by analyzing the end-to-end beatnote signal of a very narrow linewidth laser. This measurements show that the Interxion1-Telehouse2 link segment is the noisiest one.

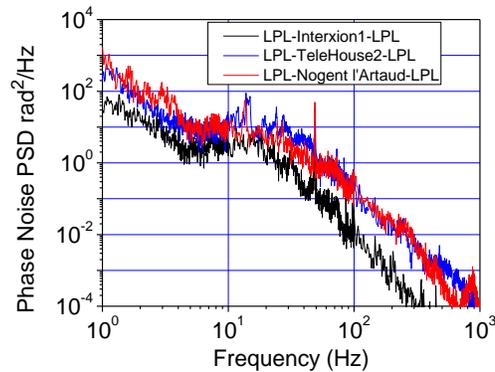

Fig. 5. Phase noise power spectral density of the different subsections of the link. Black line: first subsection LPL-Interxion1-LPL of 22 km. Blue line: LPL-Telehouse2-LPL link of 94 km. Red line: LPL-Nogent l'Artaud-LPL link of 300-km.

The optical phase noise power spectral density of the cascaded optical link of 2 x 150 km is shown in Fig. 6, without and with compensation. The phase noise reduction is around 50 dB at 1 Hz, which is very close to the theoretical limit of 51 dB [13]. For Fourier frequencies below 5 Hz, both spectra exhibit some peaks which are likely to be due to seismic noise since optical fibers on the long-haul span are buried along the railway tracks.

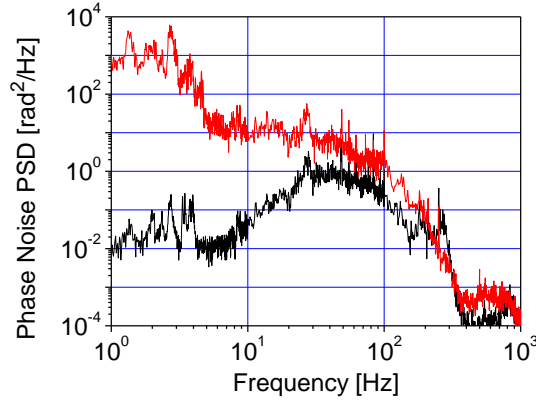

Fig. 6. Phase noise power spectral density (PSD) of the compensated 2×150 km link (black line) and twice the phase noise PSD of the free running 150 km up-link (red line).

Fig. 7 shows the fractional frequency instability (Allan deviation) of the 2×150 km link for four days of continuous operation, measured with a Π-type frequency counter. The free-running fiber frequency noise of the UL (red circles) is measured simultaneously using the compensation signal applied to the AOM located in the LPL. The Allan deviation is $3\times10^{-15}$ at 1 s averaging time and scales down as $1/\tau$ from 1 s to 70000 s reaching $5\times10^{-20}$ in a measurement bandwidth of 10 Hz (black squares). With the full bandwidth (about 100 kHz), the Allan deviation is 6 times larger (blue up-triangles).

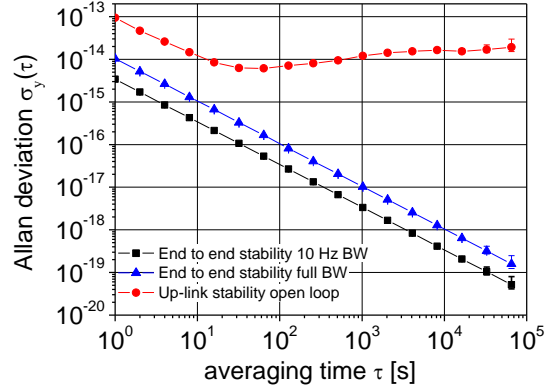

Fig. 7. End-to-end fractional frequency instability of the 150 km free running up-link (red circles), and 2x150 km compensated link measured without filter (blue up-triangles) and with a 10 Hz filter (black squares)

A crucial aspect of frequency dissemination is robustness and reliability. In our case, the optical fibers are located in an uncontrolled environment without any particular protection against thermal changes and acoustic noise. As a consequence the noise behavior is non stationary, as can be seen from the comparison between the red curves in Fig. 5 and 6, which are displaying the phase noise PSD of the 300-km link and twice the 150-km link respectively. They were recorded at different times. However the control system can easily handle these fluctuations and the link remains locked for several days. When the human activity effect, mainly in the DCF, exceeds the maximum control dynamics, phase jumps can occur with maximum amplitude of a few hundreds of ps. Those events are quite rare, at most a few per days, and induce several cycles slips without any severe unlock of the system. With long-distance propagation, the polarization state fluctuation is also a crucial point. After a few days, we had to adjust the polarization state in order to optimize the beatnote signal amplitude of the local laser phase-lock. The repeater station is not designed for an automatic adjustment of the polarization. With a single remote station, it is always possible to manually tune the polarization at the other ends of both link segments. With a multiple station link, this approach becomes impossible and the automatic polarization control is an issue.

## 4. Conclusion

We show the ultra stable transfer of an optical frequency over 300 km of installed optical fibers of an optical telecommunication network simultaneously carrying Internet data. The optical link goes through two Data Center Facilities using multiplexers and bidirectional erbium-doped fiber amplifiers. We obtained an instability of $3 \times 10^{-15}$ at 1 s which averages down to around $5 \times 10^{-20}$ after about 20 hours. This result demonstrates that non-dedicated fiber links are a valuable alternative to dark fiber links. With such Internet links, high performance frequency dissemination could be foreseen for a large ensemble of scientific users using NRENs dense fiber network facilities. In conjunction with the recently proposed PTB technique [24] to provide several users with one single optical link, optical links are now very attractive for widescale frequency dissemination. This will enable a broad range of high-sensitivity measurements, including the search for fundamental constants variation [6-8, 25] and gravitational mapping.

**Acknowledgments**

The authors are deeply grateful to D. Vandromme, T. Bono and E. Camisard from GIP RENATER and J. F. Florence from Université Paris 13 for their support in using the Renater network, and to F. Wiotte and A. Kaladjian for technical support. We acknowledge funding support from the Agence Nationale de la Recherche (ANR BLAN06-3_144016) and Université Paris 13.